\documentstyle[12pt,epsfig]{article}
\newcommand{\be}[1]{
\begin{eqnarray}\label{#1}}
\newcommand{\ee}{\end{eqnarray}}
\newcommand{\ci}[1]{\cite{#1}}
\newcommand{\re}[1]{(\ref{#1})}
\newcommand\pbar{\bar{\psi}}
\newcommand\p{\psi }

\def\Gtilde{\tilde{G}}

\newcommand{\spin}[1]{ \langle\mskip-3mu \langle{#1}
\rangle\mskip-3mu\rangle}
\newcommand{\Dperp}{\Delta_\perp}

\newcommand{\Eperpik}{\epsilon^\perp_{i k}}
\newcommand{\dxi}{\frac{\partial}{\partial \xi}}
\newcommand{\ddxi}{\frac{\partial^2}{\partial \xi^2}}
\newcommand{\du}{\frac{\partial}{\partial u}}
\newcommand{\eps}{\varepsilon}

\newcommand{\insertfig}[2]{\mbox{\epsfxsize=#1cm \epsfbox{#2.eps}}}

\begin{document}
\renewcommand{\thefootnote}{\fnsymbol{footnote}}
\begin{flushright}
\begin{tabular}{l}
 TPR-01-06\\
hep-ph/0106329\\
\end{tabular}
\end{flushright}
\begin{center}
{\bf\Large  Twist-4 
  photon helicity-flip amplitude in DVCS on a nucleon in the Wandzura-Wilczek
  approximation }

\vspace{0.5cm} N. Kivel$^{a}$\footnote{on leave of absence from 
St.Petersburg
Nuclear Physics Institute, 188350, Gatchina, Russia }, 
L. Mankiewicz$^{b,c}$, 

\begin{center}
{\em $^a$ Institut f\"ur Theoretische Physik, Universit\"at
Regensburg \\ D-93040 Regensburg, Germany}
\\ 
{\em $^b$ N. Copernicus Astronomical Center, ul. Bartycka 18,
PL--00-716 Warsaw, Poland}
\\
{\em $^c$ Andrzej Soltan Institute for Nuclear Studies,
Hoza 69, 00-689 Warsaw, Poland}

\end{center}

\end{center}

\begin{abstract}
We computed twist-4 part of the photon spin-flip amplitude in deeply 
virtual Compton scattering on a nucleon  
in the Wandzura-Wilczek
approximation. We found a factorizable contribution, which arises from photon 
scattering on quarks with non-zero angular momentum along the collision axis.
As the genuine
twist-2 amplitude arises at the NLO, for moderate virtualities of the
hard photon, $Q^2 \le 10$ GeV$^2$, kinematical twist-4
correction can give numerically important contribution to the photon
helicity-flip amplitude.
\end{abstract}

\newpage
\section*{\normalsize \bf Introduction}

Deeply virtual Compton scattering (DVCS) \cite{DVCS1,DVCS2} on a nucleon,
$\gamma^* N \rightarrow \gamma N'$ , is perhaps the
cleanest hard reaction sensitive to the skewed parton distributions
(SPD). For that reason in recent years DVCS has been the subject of extensive
theoretical investigations. First experimental data have also became recently
available (see e.g. \cite{exp1,exp2,HermesSSA, amarian}) and much more data 
are expected
from JLAB, DESY, and CERN in the near future. Due to factorization theorems
\cite{Rad97,Ji98,Col99} the leading term 
in the $1/Q^2$ expansion of
the DVCS
amplitude, where $Q^2$ is
large virtuality of the hard photon, can be
expressed in terms of twist-2 skewed parton distributions. 
However, as the
typical experimentally accessible values of $Q^2$ are by no means large,
studies of the 
power suppressed (higher twist) corrections to the DVCS amplitude are very
important from the phenomenological point of view. The
leading power corrections 
are of the order $1/Q$, or
twist-3, and therefore they may have significant effects on some of DVCS
observables.
Note also that twist-3 corrections typically scale as $\sqrt{-t}/Q$, with
$t$ denoting the square of the momentum transfer, so the size of twist-3
corrections increases with $t$. As it follows, taking into account these
corrections is mandatory for understanding continuation of the twist-2
part of the DVCS amplitude to $t=0$.

An interesting feature of the DVCS amplitude on a nucleon is that it receives
a contribution 
from 
the photon helicity-flip process, which is forbidden by the angular momentum
conservation in the forward DIS case. In the leading twist approximation, this
amplitude appears at the NLO level and, if measured, can provide a unique
information 
about tensor gluon skewed parton distribution in a nucleon. In this case,
analyzing corresponding power
corrections is even more important as there is no 
prejudice about how large the twist-2 amplitude can be. The simplest
estimate can be obtained by calculating the so-called Wandzura-Wilczek, or
kinematical power 
corrections. From a phenomenological point of view
such a calculation is crucial for the future
studies of the photon helicity-flip amplitude in DVCS.

The remainder of this paper is organized as follows: in the next section we
discuss general features of the DVCS amplitude on the nucleon. The following
two sections are devoted to discussion of the photon helicity-flip
amplitude and to the calculation of the kinematical twist four correction, 
respectively. Finally, we conclude. Technical details of the present
calculation are summarized in the Appendix.  

\section*{\normalsize \bf DVCS amplitude on a nucleon }

Let $p, p'$ and $q,q'$ denote
momenta of the initial and final nucleons and photons, respectively.
The amplitude of the virtual Compton scattering process
\be{proc}
\gamma^*(q)+N(p)\to \gamma(q') +N(p') \, ,
\ee
is defined in
terms of the nucleon matrix element of the $T$-product of two
electromagnetic currents~:
\be{T:def}
T^{\mu\nu}=-i\int d^4x\ e^{-i (q+q')x/2)}\langle p'|T\left[J_{\rm
e.m.}^\mu (x/2) J_{\rm e.m.}^\nu(-x/2)\right]|p\rangle\, ,
\ee
where Lorentz indices $\mu$ and $\nu$ correspond to the virtual, respectively 
the real
photon.

We shall consider the Bjorken limit, where $-q^2=Q^2\to\infty$, 
$2(p\cdot q)\to \infty$, with $x_B = Q^2 / 2(p\cdot q)$ constant, 
and $t \equiv (p-p')^2\ll Q^2$. 
We introduce two light-like vectors $n,\, n^*$ such that 
\be{nnstar} n\cdot n=0, \, n^*\cdot
n^*=0, n\cdot n^*=1. 
\ee  
We shall work in a reference frame where the average nucleon
momenta $P =\frac12(p+p')$ and the virtual photon momentum  $q$
are collinear along z-axis and have opposite directions.
Such a choice of the frame results in the following decomposition of the 
momenta \ci{GV}:
\be{kinem}
P&=& n^* + \frac{{\bar m}^2}{2} n \, \nonumber 
\\[4mm] 
q&=& -2\xi^\prime n^* +\frac{Q^2}{4\xi^\prime}n \, \nonumber 
\\[4mm]
\Delta&=& p'-p = - 2 \xi n^* + {\bar m}^2 \xi n + \Delta_\perp
\ee
with ${\bar m}^2 = m^2 - t/4$, $t = \Delta^2$ being the squared momentum
transfer, and 
\be{xi}
2 \xi = 2 \xi^\prime \frac{Q^2-t}{Q^2 + 4 \xi^{\prime 2} {\bar m}^2} \, .
\ee
Finally, $\xi^\prime$ is given by
\begin{equation}
\label{xiprime}
\xi^\prime 
= \frac{2}{\frac{2-x_B}{x_B} + t/Q^2 + \sqrt{(\frac{2-x_B}{x_B} + t/Q^2)^2
    + 16 \frac{ {\bar m}^2}{Q^2}}} = \frac{x_B}{2 - x_B} + O(1/Q^2) \, ,
\end{equation}
with $x_B = \frac{Q^2}{2 p \cdot q}$.

We define the transverse metric and
antisymmetric transverse epsilon tensors
\footnote{The Levi-Civita
tensor $\epsilon_{\mu \nu \alpha\beta}$ is defined as the totally
antisymmetric tensor with $\epsilon_{0123}=1$ }:
\be{gt} g^{\mu
\nu}_\perp = g^{\mu \nu}- n^\mu n^{*\, \nu}-n^\nu n^{*\, \mu},
\quad \epsilon^\perp_{\mu \nu}= \epsilon_{\mu \nu
\alpha\beta}n^\alpha n^{*\,\beta}\, . 
\ee 
In the following, we shall use the shorthand notation for 
\be{defdot}
 a^+\equiv a_\mu n^\mu, \quad  a^-\equiv a_\mu n^{*\, \mu}\, ,
\ee 
where $a_\mu$ is an arbitrary Lorentz vector.

In the LO approximation in the QCD coupling $\alpha_S$, but including $1/Q$ 
corrections, 
the DVCS amplitude on a nucleon  
has the form \cite{Penttinen,BM}~: 
\be{T}
T^{ \mu \nu}&=& T^{ \mu \nu}_1+  T^{ \mu \nu}_2+ T^{ \mu \nu}_3,
\ee
\be{T1}
T^{ \mu \nu}_1=-\frac12 \int_{-1}^1 dx\, \biggl\{ \left[g^{\mu
\nu}_\perp+{\scriptstyle \frac{P^\nu\Delta_\perp^\mu}{(Pq)} } \right]
 n^\rho F_\rho (x,\xi) C^+(x,\xi)
-\left[g^{\nu
\alpha}_\perp+{\scriptstyle \frac{P^\nu\Delta_\perp^\alpha}{(Pq)}}  \right]
i\epsilon^{\perp \mu}_\alpha n^\rho \widetilde F_\rho(x,\xi) C^-(x,\xi)
\biggl\} \nonumber \\
\ee
\be{T2}
T^{ \mu \nu}_2=
\frac{(q+4\xi P)^\mu}{(Pq)} \left[g^{\nu
\alpha}_\perp+\frac{P^\nu\Delta_\perp^\alpha}{(Pq)}  \right] 
\frac12\int_{-1}^1 dx\,\left\{
F_\alpha(x,\xi)C^+(x,\xi)- i\epsilon^\perp_{\alpha \rho}\widetilde F^\rho
(x,\xi) C^-(x,\xi)\right\}  
\ee
\be{T3}
T^{ \mu_\perp \nu}_3= \frac{(q+2\xi
P)^\nu}{(Pq)} \frac12\int_{-1}^1 dx\,
\left\{ F^{\mu_\perp}(x,\xi)
C^+(x,\xi)+i\epsilon_\perp^{\mu \rho}\widetilde F_\rho (x,\xi)
C^-(x,\xi) \right\}
\ee
where to the twist-3 accuracy: 
\be{lce} 
P&=&\frac12(p+p')=n^*, \quad \Delta = p'-p =-2\xi
P+\Delta_\perp, \, \nonumber 
\\[4mm] 
q&=&-2\xi
P+\frac{Q^2}{4\xi}n, \quad
q'=q-\Delta=\frac{Q^2}{4\xi}n-\Delta_\perp\, ,
\ee
with $\xi$ equal to its leading-order value $\xi = \frac{x_B}{2 - x_B}$.
The leading order coefficient functions are
\be{alf} 
\nonumber
C^\pm(x,\xi)=\frac{1}{x-\xi+i\varepsilon}\pm
\frac{1}{x+\xi-i\varepsilon}\, , 
\ee
and skewed distributions
$F_\mu(x,\xi)$ and  $\widetilde F_\mu(x,\xi)$  
are defined in terms
of the nonlocal light-cone quark operators \footnote{The gauge
link between points on the light-cone is not shown but always
assumed.}: 
\be{Fdef}
F_\mu(x,\xi)=\int^\infty_{-\infty}
\frac{d\lambda}{2\pi}e^{-ix\lambda} \langle p'|
\pbar(\frac12\lambda n)\gamma_\mu\p (-\frac12\lambda n)
 |p\rangle \, ,
\ee
\be{Ftlddef}
\widetilde
F_\mu(x,\xi)=
\int^\infty_{-\infty}\frac{d\lambda}{2\pi}e^{-ix\lambda} \langle p'|
\pbar(\frac12\lambda n)\gamma_\mu\gamma_5\p (-\frac12\lambda n)
 |p\rangle \, .
\ee 

In the above expression for the DVCS amplitude the first term 
$T_1^{\mu\nu}$ corresponds to the scattering of transversely
polarized virtual photons. 
This part of the amplitude depends
only on twist-2 SPDs $H,E$ and $\widetilde H, \widetilde E$.
The twist-3 terms proportional 
to  $\frac{P^\mu\Delta_\perp^\nu}{(Pq)}$
\footnote{We adopt here kinematical definition of twist i.e., 
terms suppressed by $1/Q$ are of twist-3.}
are required to ensure the proper electromagnetic gauge-invariance of the 
amplitude: 
\be{Ginv1}
q_\mu T^{\mu\nu}_1= T^{\mu\nu}_1 q'_\nu = 0\, .
\ee

The second term $T_2^{\mu\nu}$ corresponds to
the contribution of the longitudinal polarization of the virtual
photon. This term depends only
on new `transverse' SPDs $F^{\mu_\perp}$ and $\widetilde F^{\mu_\perp}$. 
Defining the longitudinal polarization vector of the virtual photon as
\be{epsL}
\varepsilon_L^\mu(q)=\frac{1}{Q}\biggl(
2\xi P^\mu+\frac{Q}{4\xi} n^\mu
\biggr)\, ,
\ee
one can easily calculate the DVCS amplitude for the longitudinal
polarization of the virtual photon ($L\to T$ transition),
which is purely of twist-3~:
\be{LtoT}
\varepsilon_\mu^L \, T^{\mu\nu_\perp}=\frac{2\xi}{Q}\int_{-1}^1dx
\ \Biggl(F^{\nu_\perp}\  C^+(x,\xi)- i\varepsilon_\perp^{\nu_\perp \alpha}
\widetilde F_\alpha\  C^-(x,\xi)
\Biggr)\, .
\ee
The skewed parton distributions
$F_\mu$ and $\widetilde F_\mu$ can be related to the twist-2 SPDs
$H,E,\widetilde H$ and $\widetilde E$ through the so-called
Wandzura-Wilczek relations \cite{BM,KPST,RW1,Kivel:2000cn,
Anikin:2001ge,Radyushkin:2001fc}. 
To derive these relations one assumes
that non-forward nucleon matrix elements of gauge invariant operators
of the type $\bar \psi G \psi$ i.e. involving quark-gluon correlations are 
small.  
The WW relations for the
case of the nucleon SPDs have the form
\cite{BM,Kivel:2000cn}~:
\be{F}
F^{WW}_\mu(x,\xi)&=&  \frac{\Delta_\mu}{2\xi}\spin{\frac
1 m}E(x,\xi)- \frac{\Delta_\mu}{2\xi}\spin{\gamma_+}(H+E)(x,\xi)+
\nonumber\\&& +\int_{-1}^{1}du\
G_\mu(u,\xi)W_{+}(x,u,\xi)+i\epsilon_{\perp \mu \alpha}
\int_{-1}^{1}du\  \widetilde G^\alpha (u,\xi)W_{-}(x,u,\xi)\, , \\
[4mm] 
\widetilde F^{WW}_\mu(x,\xi)&=&
\Delta_\mu\frac12 \spin{\frac{\gamma_5}{m}}\widetilde E(x,\xi)-
\frac{\Delta_\mu}{2\xi}\spin{\gamma_+\gamma_5}\widetilde H(x,\xi)+
\nonumber\\ && +\int_{-1}^{1}du\ \widetilde
G_\mu(u,\xi)W_{+}(x,u,\xi)+i \epsilon_{\perp \mu \alpha}
\int_{-1}^{1}du\  G^\alpha (u,\xi)W_{-}(x,u,\xi) \, .
\label{tildaF}
\ee
Here we have introduced a shorthand notation
$\spin{\ldots}=\bar U(p')\ldots U(p)$ and $m$ denotes the nucleon mass.
Functions $G^\mu$ and $\widetilde G^\mu$ are defined as~:
\be{G}
G^\mu(u,\xi)&=& \spin{\gamma^\mu_\perp}(H+E)(u,\xi)+
\frac{\Delta_\perp^\mu}{2\xi} \spin{\frac 1
m}\biggl[u\frac{\partial}{\partial u}+ \xi\frac{\partial}{\partial
\xi} \biggl] E(u,\xi)- \nonumber \\[4mm]&&
-\frac{\Delta_\perp^\mu}{2\xi}
\spin{\gamma_+}\biggl[u\frac{\partial}{\partial u}+
\xi\frac{\partial}{\partial \xi}\biggl] (H+E)(u,\xi) \, ,
\ee
\be{tG}
\widetilde G^\mu (u,\xi)& =&\spin{\gamma^\mu_\perp\gamma_5} \widetilde H(u,\xi)
+\frac12\Delta_\perp^\mu \spin{\frac{\gamma_5}{m}}
\biggl[1+u\frac{\partial}{\partial u}+\xi\frac{\partial}{\partial
\xi}\biggl] \widetilde E(u,\xi)- \nonumber\\[4mm]&&
-\frac{\Delta_\perp^\mu}{2\xi}\spin{\gamma_+\gamma_5}
\biggl[u\frac{\partial}{\partial u}+\xi\frac{\partial}{\partial
\xi}\biggl] \widetilde H(u,\xi) \, .
\ee
\indent
The Wandzura-Wilczek kernels $W_{\pm}(x,u,\xi)$
have been introduced in Refs.~\cite{BM,RW1,Kivel:2000cn}. They are
defined as~:
\be{Wpm}
W_{\pm}(x,u,\xi)&=& \frac12\biggl\{
\theta(x>\xi)\frac{\theta(u>x)}{u-\xi}-
\theta(x<\xi)\frac{\theta(u<x)}{u-\xi} \biggl\} \nonumber
\\[4mm]&&\mskip-10mu \pm\frac12\biggl\{
\theta(x>-\xi)\frac{\theta(u>x)}{u+\xi}-
\theta(x<-\xi)\frac{\theta(u<x)}{u+\xi} \biggl\}.
\ee
The flavor dependence in the amplitude can be easily restored by a
substitution~:
\be{flav}
F_\mu \, (\widetilde F_\mu) \to
\sum_{q=u,d,s, \dots}e_q^2\ F_\mu^q \, (\widetilde F_\mu^q) \, .
\ee
\indent
The amplitude \re{T} is electromagnetically gauge invariant, i.e.
\be{Ginv} q_\mu T^{\mu\nu}= T^{\mu\nu}(q-\Delta)_\nu = T^{\mu\nu} q'_\nu = 0\, ,
\ee
formally to the accuracy $1/Q^2$. In order to work with an amplitude which 
is transverse in the sense of (\ref{Ginv}) we have kept 
in \re{T2} terms of the
$\Delta^2/Q^2$ order, applying the prescription of
\cite{GV,VGG}~:
\be{transvers}
g^{\mu\nu}_\perp \, \to \, g^{\mu\nu}_\perp \,+\,
\frac{P^\nu\Delta_\perp^\mu}{(P \cdot q')} \, ,
\ee
for the twist-3 terms in the amplitude. Although such corrections do not 
form a complete set of $1/Q^2$
contributions, we prefer to work with the DVCS amplitude $T_2^{\mu\nu}$  which
satisfies Eq.~(\ref{Ginv}) exactly.

The last term $T_3^{\mu\nu}$,
corresponds to transverse polarisation of the virtual photon. It is
proportional to $(q+2\xi P)=q'+\Delta_\perp$. 
Contracting with the transverse polarisation vector $e_\nu(q')$ of
the  final real photon one obtains 
\be{zero} 
   e_\nu(q')(q+2\xi
P)^\nu=e_\nu(q')\Delta_\perp^\nu\, . 
\ee
As it follows, such term does not contribute to any observable
with the accuracy $O(\Delta/Q)$.  

In the case of a pion target, 
Ref.~\ci{RW1}, it has been shown that the
structure  $(q+2\xi P)$ emerges as truncated to the $1/Q$ accuracy
vector $q'$. 
Obviously, such a term, although formally present in the amplitude
T$^{\mu\nu}$,  
does not contribute to any physical DVCS amplitude with the the real,
transverse photon in the final state. The observation that amplitude
$T_3^{\mu\nu}$  
has zero projection onto physical states when the final photon is real is not
unexpected.  
Considering situation where both photons are virtual  
one finds that the amplitude $T_3^{\mu\nu}$ describes a $T \to L$ transition
i.e., with  
incoming transverse  and outgoing longitudinal photon, respectively. 
It has therefore to disappear in the limit when outgoing
photon is real. Indeed, the contraction
$\varepsilon_{T\mu}(q)T_3^{\mu\nu}\varepsilon_{L\nu}^* (q')$ vanishes when
$q'^2 \to 0$. The same situation is expected, of course, for a
nucleon taget.

There is another amplitude which appears at a twist-2 level,
but at the $\alpha_S$ order. This amplitude
describes DVCS of transversely polarised photons with helicity flip
between the initial and final photon states, respectively. 
Kinematical twist-4 corrections to this amplitude are the main subject 
of our considerations in this paper. At the twist-2 level the helicity-flip
amplitude depends  
on a new distribution function, so-called tensor-gluon or gluon
transversity skewed parton distribution.   
Feynman diagrams involving photon helicity flip contribution to DVCS
have been calculated by two groups \ci{BelMu,JiH}. Their result can be
represented as:
\be{Tflip}
A_{ \mu \nu}^{tw2} =
(\sum_f e^2_f)\frac{\alpha_s(Q^2)}{4\pi\xi}
\int_{-1}^1 dx\,  F_{(\mu\nu)}(x,\xi)C^-(x,\xi)\, ,
\ee
where  $F_{(\mu\nu)}$ is defined  as a matrix element of a
non-local light-cone gluon operator: 
\be{FGdef}
F_{(\mu\nu)}(x,\xi)=
\int^\infty_{-\infty}\frac{d\lambda}{2\pi}e^{-ix\lambda} \langle p'|
{\bf S}G^a_{n\mu}(\frac12\lambda n) G^a_{n\nu}(-\frac12\lambda n)
 |p\rangle \, ,
\ee
Symbols ${\bf S}$ and $(\mu\nu)$ stand for  symmetrisation of the two 
indices and removal of the trace: 
${\bf S}O_{\mu\nu}=\frac12O_{\mu\nu}+\frac12O_{\nu\mu}-
\frac14g_{\mu\nu}O^\alpha_{~\alpha}$ and $G^a_{n\nu}$ is a shorthand notation
for $n^\alpha G^a_{\alpha\nu}$.

Parametrisation of this matrix element in terms of independent SPDs 
will be discussed below. Here we note that the gluon operator \re{FGdef}
does not mix with quark operators and therefore the corresponding
SPDs  
are sometimes considered to be the cleanest probes of the gluonic content
of hadrons.   
The off-forward gluon helicity-flip SPDs in a nucleon, determined by the
matrix element of a twist-2 
operator \re{FGdef}, can be numerically as large
as other gluon
distributions. However, the
amplitude $A_{\mu\nu}^{tw2}$ arises at the NLO level
i.e. it is proportional to $\alpha_S(Q^ 2)$.  For realistic values of $Q^2$ of
order of a few GeV$^2$ it is therefore natural to consider also
power-suppressed 
corrections to the photon helicity-flip DVCS amplitude. In this paper we
provide an estimate of such higher-twist effects 
by explicitly calculating a corresponding Wandzura-Wilczek (WW) contribution
which originates from  the handbag diagram.
Note that although the WW contribution is suppressed like
$1/Q^2$, it appears already at the tree-level. 
As we shall demonstrate in the 
following, a ratio of twist-2 to twist-4
amplitudes behaves then like $\frac{\alpha_s(Q^2)}{\pi}:
\frac{m^2}{Q^2}$, which is not 
necessarily very large if $Q^2$ is of order of a few GeV$^2$. 

Recently, a similar analysis of WW contribution 
have been carried out for the
process $\gamma^*\gamma\rightarrow f_2(1270)$ \ci{KB}. 
The analog of helicity-flip amplitude in this process is the amplitude 
which describes scattering of transverse photons with different
helicities. It was found that the latter amplitude is rather sensitive to 
power corrections in the region of $Q^2 \le 10$ GeV$^2$.

The WW kinematical correction is of course not the full answer, as far as
power-suppressed corrections are concerned. Assuming that factorizability holds
for this amplitude to $1/Q^2$ accuracy, there will be additional contributions
from multiparton operators of twist four, which we have not calculated
here. Note, however, that in the chiral limit helicity is conserved
along the quark line. As it follows, in order to account for two units of
angular momentum 
one has to consider emission of two additional, transverse gluons. Current
phenomenology of power corrections is consistent with a conjecture that matrix
elements of such four-parton operators 
in a nucleon are small \ci{g2izmer}. 
This observation suggests that the WW power
correction discussed in this paper can provide a rather accurate numerical
description of  higher-twist corrections to the photon helicity-flip amplitude.

\section*{\normalsize \bf Photon helicity-flip amplitude in DVCS}

As it has been discussed in \ci{Diehl:1997bu} the LO, handbag diagram
contribution to DVCS leads to an effective s-channel photon helicity
conservation for the leading-twist amplitude. At the NLO
a new twist-2 amplitude $A_{\mu\nu}^{tw2}$ arises, which describes 
a DVCS process with photon helicity flip.

The twist-2 photon helicity-flip amplitude is absent in the handbag diagram
because of conservation of the angular momentum along the
photon-parton collision axis. As photon is a
vector particle, to allow for flip of its helicity one has to
compensate for 
two units of angular momentum. For the collinear twist-2 partonic
amplitude it is only possible by a simultaneous flip of gluon helicities, see
Fig. 1. As quarks have spin $1/2$, their helicity flip can provide at most one
unit of angular momentum. As a consequence, twist-2 photon helicity-flip 
amplitude is sensitive
to the helicity-flip gluon distribution in a nucleon. 
\begin{figure}[t]
\begin{center}
\hspace{0cm}
\insertfig{3}{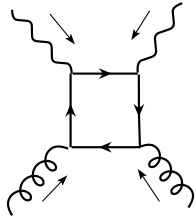}
\end{center}
\caption[dummy]{
\small  Typical diagram for the photon-gluon scattering with photon 
helicity-flip. 
Arrows indicate polarizations of photons and gluons, respectively.
}
\end{figure}
However, a similar
angular momentum conservation
argument  shows that
such distribution is forbidden in the forward limit, i.e. in DIS, on a
spin $1/2$ target. In the off-forward case a transverse
component of the 
momentum transfer $\Delta_\perp$ can provide one unit of angular momentum, so
DVCS offers the unique opportunity to investigate the  helicity-flip gluon 
distribution in a  
nucleon. As discussed in details in \ci{BelMu, JiH, Diehl:1997bu}
 such information can be 
extracted from asimuthal asymmetries $\propto \cos{3 \phi}$ of the 
cross-section, where $\phi$
is the angle between lepton and nucleon planes. 

Parametrisation of the twist-2 gluonic matrix 
element \re{FGdef} was first introduced in \ci{JiH} and recently revised in
\ci{Diehl}. 
In the notation of \ci{Diehl} one has:
\be{FGpar}
F^{(\mu\nu)}(x,\xi)&=& -{\bf S}\,\, \frac{\Delta_\perp^\mu}{4m}
\biggl\{ 
H^g_T(x,\xi) \spin{i\sigma^{+\nu}}+
\tilde H^g_T(x,\xi)\frac{\Delta_\perp^\nu}{m}\spin{1/m}\, 
\nonumber\\[4mm]&&+
E^g_T(x,\xi)\left(
\frac{\Delta_\perp^\nu}{2m}\spin{\gamma^+}+
\frac1m\spin{\gamma^{\nu}_\perp}\right)+
\tilde E^g_T(x,\xi)\frac1m\spin{\gamma^{\nu}_\perp}
\biggr\}
\ee 
Note that $\mu$ and $\nu$ in \re{FGpar} are understood as transverse Lorentz 
indices.    
As a consequence, twist-2 helicity flip amplitude \re{Tflip} is gauge 
invariant  to the $1/Q$ accuracy:
\be{Aginv}
q^\mu A_{\mu\nu}=0, \quad   A_{\mu\nu} \, q'^\nu =O(1/Q) 
\ee
Similar situation has been ecountered before in the case 
of the amplitude $T_1^{\mu\nu}$.
To get fully gauge-invariant result one has to include in the calculation 
the kinematical 
twist-3 contribution to the helicity-flip amplitude.  
As such term arises
at the NLO and
is not related to the contribution from the handbag diagram
which we are going to discuss here, it will not be considered further. 
Instead, we shall use prescription
\re{transvers} in order to obtain gauge-invariant expression. As a consequence, combining 
twist-2 and 
twist-4 contributions, one can write  the helicity-flip 
amplitude $A^{\mu\nu}$ in the following form:
\be{T2dec}
A^{\mu\nu}=\frac12\left[g^{\mu i}_\perp
\left(g^{\nu j}_\perp+{\scriptstyle 
\frac{P^\nu\Delta_\perp^j}{(Pq)}} \right)+
g^{\mu j}_\perp 
\left(g^{\nu i}_\perp+{\scriptstyle
\frac{P^\nu\Delta_\perp^i}{(Pq)}}\right)-
\left(g^{\mu \nu}_\perp+{\scriptstyle
\frac{P^\nu\Delta_\perp^\mu}{(Pq)}} \right)
g^{ij}_\perp\right]
\biggl[
A_{(ij)}^{tw2}+\frac{m^2}{Q^2}A_{(ij)}^{tw4}
\biggl] \, .
\ee
Here $A_{(ij)}^{tw2}$ is the leading twist contribution \re{Tflip}. 
The twist-4 part, arising from the handbag
diagram through the WW mechanism can be parametrized as 
$\frac{m^2}{Q^2}A_{(ij)}^{tw4}$, with $m$ being the nucleon mass. A convenient 
physical interpretation of this term follows from an observation that
beyond the leading  
twist approximation one can imagine partons as carrying
non-zero orbital angular momentum along the collision axis. That allows quarks
to participate in the LO, i.e. through the handbag diagram, in
the photon helicity-flip amplitude. As two units of angular momentum have to
flow through the hard vertex, such an amplitude is suppressed by $1/Q^2$ and
represents a twist-4 contribution.

Following \ci{Diehl}, one observes that the twist-2 gluonic matrix element 
\re{FGdef} can be parametrized in terms of  four independent
SPD:  $H^g_T,\, E^g_T,\, \tilde H^g_T$ and $\tilde E^g_T$ associated with four 
transverse tensor structures: 
\be{cheven}
\mbox{chiral even:  }\quad {\bf S}\,\frac1{m^2}
\Delta_\perp^i\spin{\gamma^{j}_\perp} , \quad
 {\bf S}\, \frac{\Delta_\perp^i\Delta_\perp^j}{m^2}\spin{\gamma^+}, \,
\\[4mm]
\mbox{chiral odd:}\quad
{\bf S}\, \frac1m 
\Delta_\perp^i\spin{i\sigma^{j+}}, \quad
{\bf S}\, \frac{\Delta_\perp^i\Delta_\perp^j}{m^2}\spin{1/m}.
\label{chodd}
\ee
Here, four independent structures correspond to four independent helicity-flip
amplitudes in the gluon-nucleon system. As the number of independent 
quark-nucleon helicity-flip amplitudes is the same
\footnote{Note that we consider here
only amplitudes with photon helicity flip by two units.}, 
one expects that these tensor structures will appear 
in the twist-4 amplitude calculated in the WW approximation as well. 
As it follows, the same basis of Dirac structures (\ref{cheven},\ref{chodd})
can be used as the basis for an expansion of
$A_{(ij)}^{tw4}$.

\section*{\normalsize \bf Photon helicity-flip amplitude in the
Wandzura-Wilczek approximation}

Let us now  discuss briefly calculation of
the WW contribution  to the 
photon helicity-flip amplitude in DVCS. Formally, the WW contribution arises 
because  operators with external derivatives w.r.t. total translation in a
transverse direction 
give nonzero
contribution in the DVCS kinematics.

The photon helicity-flip amplitude \re{T2dec} is symmetric in the
indices $\mu$ and $\nu$ and therefore it arises from  
the symmetric 
part of the T-product of the electromagnetic currents 
in \re{T:def}.  
The tree-level contribution results from the handbag diagram 
depicted in Fig. 2. It can be written as
\be{HB}
T^{\mu\nu}=\frac1{\pi^2}\int d^4 x e^{-i(q+q')x}
s^{\mu\nu}_{\lambda\sigma}\frac{x_\lambda}{x^4}
\langle p'|
\pbar ( x)\gamma_\sigma \p(- x)-\pbar ( -x)\gamma_\sigma \p(x)
|p \rangle\, + \dots
\ee 
where ellipses denote the contribution antisymmetric in $\mu,\nu$, and
$s^{\mu\nu}_{\lambda\sigma} = g^\mu_\lambda g^\nu_\sigma + g^\mu_\sigma
g^\nu_\lambda - g^{\mu\nu} g_{\lambda\sigma}$.
In order to obtain the twist-4 WW contribution to the amplitude \re{T2dec}, 
we have
extracted from the
matrix element
\be{me}
\langle p'|\pbar ( x)\gamma_\sigma \p(- x)|p \rangle
\ee
terms linear and bilinear in transverse structures
$\gamma_\perp$ and $\Delta_\perp$. The linear terms can in principle
contribute because to obtain the full answer one has to expand the exponent
$e^{-i(q+q')x}$ in $\Delta_\perp$ as well. In the WW approximation one
neglects, as usual, contribution from quark-gluon operators. 
The final answer is then given in 
terms of twist-2 SPDs related to the matrix elements of vector and axial
operators  $H,\, E$ and $\widetilde H,\, \widetilde E $ 
respecively, see \re{F},\re{tildaF}. 

\begin{figure}[t]
\begin{center}
\hspace{0cm}
\insertfig{8}{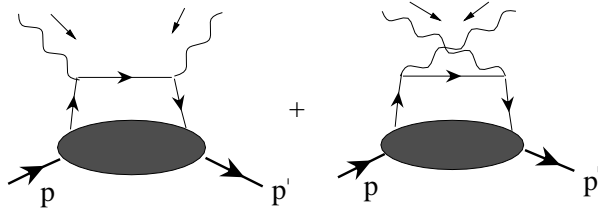}
\end{center}
\caption[dummy]{
\small Typical handbag diagrams which contribute to the twist-4 
amplitude disussed in 
the text. Arrows indicate polarization of photons. Angular-momentum 
conservation requires that quarks carry an orbital angular momentum along the 
collision axis.
}
\end{figure}

The calculation is straightforward and technically rather close 
to the approach described in \ci{Kivel:2000cn}. Technical details of the
present 
work are summarized in the 
Appendix.  For a detailed
discussion of a similar calculation, but with 
matrix elements parametrised in terms of double distributions, see Ref.  
\ci{RW1}.
Note that because in the WW approximation one neglects quark-gluon operators, 
gluon emission diagrams, calculated in 
\ci{BM,Anikin}, have not been taken into account here.

We have also found that apart from terms
which 
contribute to the photon helicity-flip amplitude \re{A4} in the WW
approximation, one finds also a singular, {\it non-factorizable} term $\sim
\Delta_\perp^\mu \Delta_\perp^\nu/(Pq)$ which contributes to the amplitude
$T_3^{\mu\nu}$ in such a way that the factor $(q + 2 \xi
P)^\nu$ there becomes equal to $q'^\nu$. Thus, the non-factorizable term gives
no contribution to the physical DVCS amplitudes with a real photon in 
the final state, as discussed in the previous section.

The explicit expression for the twist-4
amplitude defined in \re{T2dec} reads:
\be{A4}
A^{(i j)}_{tw4}&=& \xi \int_{-1}^1 dt C^-(t,\xi)\biggl\{
\frac1{m^2}\left(\Dperp^i G^j(u,\xi)+\Dperp^j G^i(u,\xi) - 
{\rm trace}\right) \otimes U_-(u,t,\xi)
+\nonumber \\[4mm]
&+&
\frac1{m^2}
\left(\Dperp^i i \epsilon_\perp^{jk}\Gtilde_k(u,\xi)+\Dperp^j 
\epsilon_\perp^{i k}\Gtilde_k(u,\xi)- {\rm trace}
\right)\otimes U_+(u,t,\xi)\biggr\}+
\\[4mm]
&+&2\xi \int_{-1}^1 dt \left[
\ln \left(t-\xi+i\epsilon \right)-\ln \left(t+\xi-i\epsilon \right)
\right]\biggl\{ 
H_{\rm S}^{i j}(u,\xi)+
\nonumber \\[4mm]
&+&G_{\rm S}^{i j}(u,\xi)
\otimes W_+(u,t,\xi)+
\frac1{2}\left(i \epsilon_\perp^{i k}\Gtilde_k^{~j}+  
i \epsilon_\perp^{j k}\Gtilde_k^{~ i}\right)(u,\xi)
\otimes W_-(u,t,\xi) - {\rm traces }\biggr\} \, . \nonumber
\ee  
This is the main result of the present paper.
For the sake of clarity, we have introduced a shorthand notation for the
convolution integrals, 
e.g. 
\be{conv}
G_k(u,\xi)\otimes U_+(u,t,\xi)\equiv \int_{-1}^1 d u G_k(u,\xi)U_+(u,t,\xi).
\ee 
The flavour structure can be restored according to \re{flav}. Note that
although the twist-2 amplitude \re{Tflip} is a flavor-singlet,
twist-4 WW correction arising from the handbag diagram is a mixture of
singlet and non-singlet flavor contributions.
All tensors which appear above are understood to have transverse Lorentz
indices.
Functions $G_k, \tilde G_k$ and kernels $W_\pm(u,t,\xi)$ are   
defined in \re{F} and their explicit form is given in \re{G},\re{tG} and
\re{Wpm}. Functions 
$H^{i j}_S,\, G^{i j}_S,\, \tilde G^{i j}$ and 
kernels $U_\pm(u,t,\xi)$ are defined below.

Explicit expressions for $U_\pm(u,t,\xi)$ read:
\be{Ukern}
U_{\pm}(u,x,\xi)&=& \frac12(x-\xi)\biggl\{
\theta(x>\xi)\frac{\theta(u>x)}{(u-\xi)^2}-
\theta(x<\xi)\frac{\theta(u<x)}{(u-\xi)^2} \biggl\} \nonumber
\\[4mm]&&\mskip-10mu \pm\frac12(x+\xi)\biggl\{
\theta(x>-\xi)\frac{\theta(u>x)}{(u+\xi)^2}-
\theta(x<-\xi)\frac{\theta(u<x)}{(u+\xi)^2} \biggl\}.
\ee
To establish factorization, it is crucial to inspect properties of convolution
integrals which appear in \re{A4} at the     
points $t=\pm \xi$. For a test function
$f(u,\xi)$  one readily finds
\be{Uprop}
\lim_{\eps\rightarrow 0} f(u,\xi)\otimes \biggl[ 
U_-(u,t=\pm \xi+\eps,\xi)-U_-(u,t=\pm\xi-\eps,\xi)\biggr]=0
\nonumber \\[4mm]
\lim_{\eps\rightarrow 0} f(u,\xi)\otimes \biggl[ 
U_+(u,t=\pm\xi+\eps,\xi)-U_+(u,t=\pm\xi-\eps,\xi)\biggr]=
\nonumber \\[2mm]
= f(\pm\xi+0,\xi)-f(\pm\xi-0,\xi)\, .\mskip100mu 
\ee

One observes that the kernel $U_{+}$ appears in \re{A4} in convolution 
with function $\Gtilde_k(u,\xi)$ which is
defined in terms twist-2 SPDs
and their derivatives, see \re{tG}.
Note that although in general  derivatives of a SPD 
with respect to $x$
and $\xi$ are discontinuous at $x=\pm\xi$, the
combinations $(x\partial_x+\xi\partial_\xi)\widetilde 
E(\widetilde H)(x,\xi)$ are continuous
at these points \ci{Kivel:2000cn}. Hence, from \re{tG} it follows that   
$\Gtilde_k(u,\xi)$ has no dicontinuities at $u=\pm \xi$. 
As a result, convolution integrals which appear in the first two
terms in \re{A4} define a continuous function of $t$ at $t=\pm \xi$ and 
factorisation   
is not violated.  

The  third term in \re{A4} has only logarithmic, integrable
singularities at the points  $x=\pm \xi$ in the
coefficient function.
One concludes therefore that this convolution integral is also well 
defined. Explicit expressions for functions 
$H_{\rm S}^{i j},\, G_{\rm S}^{i j},\, 
\tilde G^{i j}$  read 
\be{Hsj}
H_{\rm S}^{i j}(u,\xi)
&=&\frac{\Dperp^i\Dperp^j}{4m^2\xi^2}
\left(1-\xi\dxi\right)\left\{ 
\spin{\frac1m}E(u,\xi)
 -\spin{\gamma^+}(H+E)(u,\xi)
\right\}
\nonumber\\[4mm]
&-&\frac{1}{2m}\left\{\frac{\Dperp^i}{2m\xi}\spin{\gamma_\perp^j}+ 
\frac{\Dperp^j}{2m\xi}\spin{\gamma_\perp^i}\right\}(H+E)(u,\xi)
\ee
\be{Gsj}
G_{\rm S}^{i j}(u,\xi)&=&
\spin{\gamma^+}\frac{\Dperp^i\Dperp^j}{4m^2\xi^2}
\left[\xi^2\ddxi- \left(1-\xi\dxi\right)u\du\right](H+E)(u,\xi)
\nonumber\\[4mm]
&-&\spin{\frac1m}\frac{\Dperp^i\Dperp^j}{4m^2\xi^2}
\left[\xi^2\ddxi- \left(1-\xi\dxi\right)u\du\right]E(u,\xi)
\\[4mm]
&-&\frac{1}{2m}\left\{\frac{\Dperp^i}{2m\xi}\spin{\gamma_\perp^j}+
\frac{\Dperp^j}{2m\xi}\spin{\gamma_\perp^i}\right\}
\left[2\xi\dxi+u\du\right](H+E)(u,\xi), \nonumber
\ee
\be{tildGsj}
\Gtilde^{i j}(u,\xi)&=& \spin{\gamma^+\gamma_5}\frac{\Dperp^i\Dperp^j}{4m^2\xi^2}
\left[\xi^2\ddxi-u\du\left(1-\xi\dxi\right)\right]\widetilde H(u,\xi)
\nonumber\\[4mm]
&-& \spin{\frac{\gamma_5}m}\frac{\Dperp^i\Dperp^j}{4m^2\xi}
\left[\xi^2\ddxi+\xi\dxi\left(2+u\du\right)\right]\tilde E(u,\xi)
\nonumber\\[4mm]
&-&\frac{1}{m}\left\{\frac{\Dperp^i}{2m\xi}\spin{\gamma_\perp^j\gamma_5}+
\frac{\Dperp^j}{2m\xi}\spin{\gamma_\perp^i\gamma_5}\right\}
\xi\dxi \widetilde H(u,\xi)
\nonumber\\[4mm]
&+&\frac{1}{m}\left\{
\frac{\Dperp^j}{2m\xi}\spin{\gamma_\perp^i\gamma_5}
-\frac{\Dperp^i}{2m\xi}\spin{\gamma_\perp^j\gamma_5}(1+u\du) 
\right\}\widetilde H(u,\xi)
\ee

Note that functions $\tilde G_k$ and $\Gtilde^{i j}$ contain
Dirac structures which are defined with the help of the  
$\gamma_5$ matrix. Using relations which follow from the Gordon
identities one can express them through the basic structures \re{cheven} and
\re{chodd}. For example, one finds
\be{spinid}
\Eperpik\Dperp^k \spin{\gamma_5}= - \frac12 t \spin{\sigma^+_{~i}}-
2 i \xi m \spin{\gamma_{\perp i}} - \frac12 i \Delta_{\perp i}(2 m
\spin{\gamma^+} - 2 \spin{1}) \, ,
\nonumber\\[4mm]
2 m \Eperpik\Dperp^k \spin{\gamma^+ \gamma_5}= - 2 \xi\Eperpik \Dperp^k
\spin{\gamma_5} - 2 i (2 m \spin{\gamma_{\perp i}} - \xi \Delta_{\perp i}
\spin{1} + 2 i \xi {\bar m}^2 \spin{\sigma^+_{~i}}) \, ,
\nonumber\\[4mm]
2 m \Eperpik \spin{\gamma_\perp^k \gamma_5} = \Eperpik \Dperp^k \spin{\gamma_5}
- i \Delta_{\perp i} \spin{1} - 2 {\bar m}^2 \spin{\sigma^+_{~i}}\, ,
\nonumber\\[4mm]
2 m \Eperpik \Dperp^k \spin{\gamma_{\perp j}\gamma_5} = \Delta_{\perp j}
\Eperpik \Dperp^k \spin{\gamma_5} - i \Delta_{\perp i} \Delta_{\perp j}
\spin{1} - 2 {\bar m}^2 \spin{\sigma^+_{~i}} + \dots \, .
\ee
Here ellipses stand for terms which are
proportional to $\delta_{ij}$ and therefore do not contribute to the traceless
combination which enters amplitude \re{A4}. Note that, as expected,  all
tensor structures present in the twist-2 matrix element \re{FGpar} appear 
also in the twist-4 WW contribution. 

It is important to stress  that all integrals which define twist-4 amplitude
$\frac{m^2}{Q^2}A^{\sigma j}_{tw4}$  
in \re{A4} are well defined and therefore factorisation is not violated, as far
as the tree-level WW contribution to the twist-4 photon helicity-flip amplitude
is concerned. This is an interesting result, as current QCD
factorization theorems for exclusive processes guarantee factorization of the
leading-twist contribution to DVCS only. 

Let us now consider in more detail contributions of the so-called
{\it pion pole} and {\it D-terms}.
At a small $t$ the skewed parton
distribution $\widetilde E$ is dominated by chiral
contribution of the pion pole \cite{pion, GPV} of the form:
\be{pipo} 
\widetilde E^{\rm pion\ pole}(x,\xi)=\frac{4 g_A^2
m^2}{-t+m_\pi^2}\ \frac 1\xi \varphi_\pi\left(\frac x\xi
\right)\theta(|x|\le\xi)\, , 
\ee 
where $g_A$ is the axial charge
of the nucleon and $\varphi_\pi(u)$ is the pion distribution
amplitude. As it was noted in \ci{Kivel:2000cn} this contribution  cancel
in $\Gtilde^{k}$. It is easy to see that
in the function $\Gtilde^{i j}$ defined in \re{tildGsj} 
pion pole contribution again vanishes under action of the 
differential operator.  It can be understood as a consequence of 
P-invariance  since a pseudoscalar t-channel exchange cannot
contribute to the photon helicity-flip amplitude. 
Vanishing of the pion
pole contribution is therefore a non-trivial check of our
calculation. 

D-terms \ci{Dterm} complete  parametrisations of
SPDs in terms of double distributions \cite{RadDD}. They have
the form: 
\be{Dterm} 
\nonumber H^{\rm
D-term}(x,\xi)&=&~D\left(\frac{x}{\xi}\right)\ \theta(|x|\le\xi)\,
,\\ E^{\rm D-term}(x,\xi)&=&-D\left(\frac{x}{\xi}\right)\
\theta(|x|\le\xi)\, . 
\ee 
Here $D(u)$ is an odd function of its
argument. Estimates in the framework of the chiral quark-soliton model of a 
nucleon suggest that D-term can be numerically large \cite{bor}.  Calculation 
of the DVCS cross-section 
shows that effects of the D-term can be clearly seen \ci{Kivel:2000cn, GPV}.
In the present case one readily finds that D-term gives non-zero contribution 
only through 
function $H_{\rm S}^{i j}(u,\xi)$:
\be{HDt}
H_{\rm S}^{i j}(u,\xi)\biggl|_{\rm D-term}=- 
\frac{\Dperp^i\Dperp^j}{4m^2\xi^2}\spin{\frac1m}\theta(|x|\le\xi)
\left[ D\left(\frac{x}{\xi}\right)+ 
\frac{x}{\xi} D'\left(\frac{x}{\xi}\right) \right]   
\ee
where $D'(x)\equiv d/dx D(x)$.
In all other functions introduced here the D-term contribution vanishes under 
action of 
differential operators. 

Detailed investigation of relative importance of the twist-4
correction as compared to the twist-2 amplitude requires models for
both gluon transversity and twist-2 skewed quarks distributions in a
nucleon. However, a qualitative analysis shows that the WW
contribution to the photon helicity-flip amplitude can be large and
its consideration is perhaps mandatory for any attempt to extract an
estimate of the gluon transversity distribution from the
data. Assuming that the ratio of convolution integrals of SPD with
corresponding Wilson coefficients is of the order of one, from
\re{FGpar} and \re{A4} one finds that the ratio of twist-2 to
twist-4 amplitudes behaves like
$\frac{\alpha_s(Q^2)}{\pi} : m^2/Q^2$, which gives numerically $\sim
0.25$ and $\sim 1.2$ for $Q^2 = 2$ and $10$ GeV$^2$,
respectively. This suggests that 
 it might be
necessary to take into account the WW contribution to the photon
helicity-flip amplitude up to values of $Q^2$ of the order of 10 GeV$^2$.

\section*{\normalsize \bf Summary and conclusions}

 It has been recognized for some time that photon helicity-flip amplitude in 
DVCS on a nucleon provides a unique opportunity to study the twist-2 
gluon transversity distribution in a nucleon. Because of its importance for 
studies of novel aspects of nucleon structure, it is mandatory to 
consider not only the leading-twist contribution, but the 
power-suppressed corrections to this amplitude as well. 

In this paper we have calculated twist-4 correction to the photon helicity-flip
amplitude in DVCS in the Wandzura-Wilczek approximation. It
originates at the LO from scattering of the virtual 
photon on quarks which carry a 
non-zero
projection of angular momentum along the collision axis. We found a 
factorizable formula which allows to calculate the kinematical power 
correction in terms of twist-2 quark skewed parton 
distributions.

Numerically, the power correction discussed in 
this paper is relatively enhanced as compared to the twist-2 amplitude 
which arises at the NLO.
As a consequence, for moderate virtualities of the
hard photon, $Q^2 \le 10$ GeV$^2$, kinematical twist-4
correction might give an important contribution to the photon
helicity-flip amplitude.

\section*{\normalsize \bf Acknowledgments}
We would like to thank   V.~Braun and M.~Polyakov for fruitful discussions.
We thank A. Radyushkin and M. Diehl for critical reading of the manuscript. We
are  
grateful to K. G\"oke for hospitality extended to us during our visit to 
Bochum University, where this work was completed.
The work of N.K. was supported by the DFG, project No.~920585.
This work has been supported in part by the KBN grant
2~P03B~011~19

\section*{\normalsize \bf APPENDIX: Twist-4 contribution 
to the matrix element of the vector quark operator }
\setcounter{equation}{0}
\label{app:a}
\renewcommand{\theequation}{A.\arabic{equation}}
\setcounter{table}{0}
\renewcommand{\thetable}{\Alph{table}}

In this Appendix we present details of the
calculation of the matrix element \re{me} in the WW approximation. 
In order to compute twist-4
correction to 
photon helicity-flip amplitude in DVCS one has
to expand \re{me} to twist-4 accuracy, retaining terms which give a
non-zero 
contribution to the symmetric, traceless amplitude $A^{\mu\nu}$. 
The resulting expansion for arbitrary $x^2\neq 0$ has the form
\be{Tdec}
\langle p'|\pbar ( x)\gamma^\sigma \p(- x)|p \rangle&=&
P^\sigma V_{\rm tw2}(Px,\xi)
+V_{\rm tw3}^{\sigma_\perp}(Px,\xi)
+
V_{\rm tw4}^{\sigma_\perp \rho_\perp}(Px,\xi) x_{\rho_\perp}
+\dots \, .
\ee
Note that $\xi=-\frac12\frac{(\Delta n)}{(P
n)}$ and ellipses denote twist-4 part
which results in a term $\sim g^{\mu\nu}_\perp$
and therefore do not
contribute to the amplitude $A^{\mu\nu}$.

The twist-3 term $V_{\rm tw3}^{\sigma_\perp}$ in the 
expansion has been obtained in
\cite{BM,Kivel:2000cn}, see Eq. \re{F}.
Calculation of the twist-4 term $V_{\rm tw4}^{\sigma_\perp
\rho_\perp}$ is equivalent to expansion of the
matrix element \re{me} up to terms bilinear in
the transverse structures
$\gamma_\perp$ and $\Delta_\perp$.  The result reads:
\be{Ttw4}
\begin{array}{lll}
\displaystyle
V^{\sigma \rho}(Px,\xi) x_\rho = 
i(x\Dperp ) 
\int_{-1}^1 dt e^{2i t(Px)}
\left[ G^\sigma(u,\xi)\otimes U_-(u,t,\xi)+ 
i\epsilon_\perp^{\sigma \rho} \tilde G_\rho(u,\xi)\otimes U_+(u,t,\xi)\right]
\\[4mm]\displaystyle
-2m^2 x_\rho\left\{
\int_{-1}^t d u \left[H^{\sigma \rho}(u,\xi)+ G^{\sigma \rho}(u,\xi)\right]+
G^{\sigma \rho}(u,\xi)\otimes
\left[tW_+(u,t,\xi)-\xi W_-(u,t,\xi)\right]\right.
\\[4mm]\displaystyle
\mskip140mu +i\epsilon_\perp^{\sigma \alpha} \tilde G_\alpha^{~ \rho}(u,\xi)
\otimes\left[tW_-(u,t,\xi)-\xi
 W_+(u,t,\xi)\right]\biggr\}
\\[4mm]\displaystyle
 - i  (x\Dperp ) \int_{-1}^1 dt e^{2i t(Px)} 
\left[
G^\sigma(u,\xi)\otimes W_-(u,t,\xi)+
i\epsilon_\perp^{\sigma \alpha} \tilde G_\alpha(u,\xi)
\otimes W_+(u,t,\xi)\right] \, .
\end{array}
\ee
Here all Lorentz indices are understood to be transverse.
The first three lines of the above expression contribute
to the photon helicity-flip amplitude. The last line
results in a divergent, {\it non-factorizable} contribution to the
amplitude $T_3^{\mu\nu}$. As discussed in the text, this contribution vanishes
when contracted with a polarization vector of the final, transverse, real
photon.

All functions appearing in the right-hand-side of the above equation, except
$G^{\sigma \rho}(u,\xi)$ and $H^{\sigma \rho}(u,\xi)$,
have been defined in the main body of the paper, see Eqs. \re{F} -- \re{tG} for
$ G_k(u,\xi),\,\tilde G_k(u,\xi)$ and $W_\pm (u,t,\xi)$ and 
\re{Ukern} and \re{tildGsj} for $U_\pm (u,t,\xi)$ and $\tilde G^{k
  \rho}(u,\xi)$,
respectively.  The remaining terms are defined as follows:
\be{G2}
G^{\sigma \rho}(u,\xi)&=&
\spin{\gamma_+}\frac{\Dperp^\sigma\Dperp^\rho}{4m^2\xi^2}
\left[\xi^2\ddxi- \left(1-\xi\dxi\right)u\du\right](H+E)(u,\xi)
\nonumber\\[4mm]
&-&\spin{\frac1m}\frac{\Dperp^\sigma\Dperp^\rho}{4m^2\xi^2}
\left[\xi^2\ddxi- \left(1-\xi\dxi\right)u\du\right]E(u,\xi)
\nonumber\\[4mm]
&-&\frac{1}{m}\left\{\frac{\Dperp^\sigma}{2m\xi}\spin{\gamma_\perp^\rho}+
\frac{\Dperp^\rho}{2m\xi}\spin{\gamma_\perp^\sigma}\right\}
\xi\dxi(H+E)(u,\xi)
 \nonumber\\[4mm]
&&-\frac{\Dperp^\sigma}{2m^2\xi}\spin{\gamma_\perp^\rho}
u\du (H+E)(u,\xi)
\ee 
\be{H2}
H^{\sigma \rho}(u,\xi)
&=&\frac{\Dperp^\sigma\Dperp^\rho}{4m^2\xi^2}
\left(1-\xi\dxi\right)\left\{ 
\spin{\frac1m}E(u,\xi)
 -\spin{\gamma_+}(H+E)(u,\xi)
\right\}
\nonumber\\[4mm]
&-&\frac{\Dperp^\sigma}{2m^2\xi}\spin{\gamma_\perp^\rho}(H+E)(u,\xi)
\ee
Note that, by comparing with Eqs. (\ref{Hsj},\ref{Gsj}), one finds:
\be{HGrel}
H^{\sigma \rho}_S(u,\xi)&=&\frac12\left(H^{\sigma \rho}(u,\xi)+
H^{\rho\sigma}(u,\xi) 
\right)\, ,
\\[4mm]
G^{\sigma \rho}_S(u,\xi)&=&\frac12\left(
 G^{\sigma \rho}(u,\xi)+G^{\rho\sigma }(u,\xi)\right) \, .
\ee

Note that one can obtain a sum rule for the twist-4 part defined in
\re{Ttw4}.   
To this end, let us consider the parametrisation of the matrix element
of the quark part of the energy momentum tensor \ci{Spin}:
\be{Qme}
\frac12 \langle p'|\pbar \frac12 \left[\gamma^\mu i\stackrel{\leftrightarrow}{D}^\nu
 + \gamma^\nu i\stackrel{\leftrightarrow}{D}^\mu \right]
\p |p \rangle= [A(\Delta^2)+B(\Delta^2)] P^{\{\mu}\spin{\gamma^{\nu\}}}
\nonumber
\\[4mm]
-P^\mu P^\nu B(\Delta^2)\spin{\frac1m}
+C(\Delta^2)(\Delta^\mu \Delta^\nu-g^{\mu\nu}\Delta^2)\spin{\frac1m}+
\bar C(\Delta^2)g^{\mu\nu}m\spin{1} .
\ee
Here $\{\mu\nu\}$ denotes symmetrisation with respect to the indices 
$\mu$ and $\nu$.
Contracting both sides of this equation with $n_\mu n_\nu$
and using
\re{F} one finds \ci{Spin}:
\be{plus}
\int_{-1}^{1}dt tH(t,\xi)= B(\Delta^2)+4\xi^2 C(\Delta^2),
\nonumber \\[4mm]
\int_{-1}^{1}dt tE(t,\xi)= B(\Delta^2)-4\xi^2 C(\Delta^2)
\ee
Note
that formfactor $C(\Delta^2)$ corresponds entirely to the contribution of the 
D-term  \ci{Dterm}. On the other hand, by taking the transverse, traceless
projection one finds with the help of \re{Ttw4}:
\be{perp}
\frac12\langle p'|\pbar \frac12 
\left[\gamma^\mu_\perp i\stackrel{\leftrightarrow}{D}^\nu_\perp
 + \gamma^\nu_\perp i\stackrel{\leftrightarrow}{D}^\mu_\perp-{\rm trace} \right]
\p |p \rangle&=& \frac14 
\left[ V^{\mu_\perp \nu_\perp}(0,\xi)+
V^{\nu_\perp \mu_\perp}(0,\xi)-{\rm trace}
\right] 
\nonumber\\[4mm]
&=&C(\Delta^2)\Dperp^{(\mu} \Dperp^{\nu)}\spin{\frac1m}.
\ee
This sum rule provides a non-trivial check of the expansion \re{Ttw4}.
Direct calculation gives: 
\be{Srule}
\frac14\left[ V^{\mu_\perp \nu_\perp}(0,\xi)+
V^{\nu_\perp \mu_\perp}(0,\xi)-{\rm trace}
\right] =  
(-1)\Dperp^{(\mu} \Dperp^{\nu)}\spin{\frac1m}\frac18
\frac{d^2}{d\xi^2}\int_{-1}^{1}dt tE(t,\xi).
\ee
Using \re{plus} one finds an agreement between \re{perp} and \re{Srule}.

Finally, for the sake of completness, let us mention two useful 
relations which allow to simplify the rhs of \re{Ttw4}.
One can easily check that the following equations hold:
\be{rel}
\frac{d}{d t}\int_{-1}^{1}du \tilde G^{\sigma \rho}(u,\xi) 
\left[tW_-(u,t,\xi)-\xi W_+(u,t,\xi)\right]=
\int_{-1}^{1}du \tilde G^{\sigma \rho}(u,\xi)W_-(u,t,\xi), 
\nonumber \\[4mm]
 \frac{d}{d t}\int_{-1}^{1}du G^{\sigma \rho}(u,\xi)
\left[tW_+(u,t,\xi)-\xi W_-(u,t,\xi)\right]=
\int_{-1}^{1}du G^{\sigma \rho}(u,\xi)W_+(u,t,\xi) - 
G^{\sigma \rho}(t,\xi).\nonumber
\ee
We have made use of these relations in order to bring the expression for the 
amplitude \re{A4} to
the form quoted in this paper.


\begin{thebibliography}{99}


\bibitem{DVCS1}
D. M\"uller, D. Robaschik, B. Geyer, F.M. Dittes, J. Horejsi,
Fortschr. Phys. {\bf 42} (1994) 101.
\bibitem{DVCS2}
X. Ji, Phys. Rev.{\bf D55} (1997) 7114.
\bibitem{exp1}
P.~R.~Saull  [ZEUS Collaboration], ``Prompt photon production and
observation of deeply virtual Compton  scattering,''
hep-ex/0003030.
\bibitem{exp2} Rainer Stamen [H1-Collaboration],
`` Measurement of the Deeply Virtual Compton Scattering at Hera'',
H1prelim-00-17, DIS 2000 and IHEP 2000.

\bibitem{HermesSSA}
A.~Airapetian  [HERMES Collaboration],
hep-ex/0106068.

\bibitem{amarian} M.~Amarian [HERMES collaboration],
``DVCS and exclusive meson production measured by HERMES'',\\ talk
at workshop ``Skewed Parton Distributions and Lepton - Nucleon
Scattering'', DESY, Sept. 2000,
http://hermes.desy.de/workshop/TALKS/talks.html

\bibitem{Rad97}
A.~V.~Radyushkin, Phys.\ Rev.\  {\bf D56} (1997), 5524.
\bibitem{Ji98}
X.~Ji and J.~Osborne, Phys.\ Rev.\  {\bf D58} (1998) 094018.
\bibitem{Col99}
J.~C.~Collins and A.~Freund, Phys.\ Rev.\  {\bf D59} (1999)
074009.

\bibitem{GV}
P.~A.~Guichon and M.~Vanderhaeghen,
Prog.\ Part.\ Nucl.\ Phys.\  {\bf 41} (1998) 125
[hep-ph/9806305].

\bibitem{Penttinen}
M.~Penttinen, M.~V.~Polyakov, A.~G.~Shuvaev and M.~Strikman,
Phys.\ Lett.\ B {\bf 491} (2000) 96
[hep-ph/0006321].


\bibitem{BM}
A.~V.~Belitsky and D.~Muller,
Nucl.\ Phys.\ B {\bf 589} (2000) 611
[hep-ph/0007031].


\bibitem{KPST}
N.~Kivel, M.~V.~Polyakov, A.~Schafer and O.~V.~Teryaev,
Phys.\ Lett.\ B {\bf 497} (2001) 73
[hep-ph/0007315].


\bibitem{RW1}
A.~V.~Radyushkin and C.~Weiss,
Phys.\ Lett.\ B {\bf 493} (2000) 332
[hep-ph/0008214].
\\
A.~V.~Radyushkin and C.~Weiss,
Phys.\ Rev.\ D {\bf 63} (2001) 114012
[hep-ph/0010296].



\bibitem{Kivel:2000cn}
N.~Kivel and M.~V.~Polyakov,
Nucl.\ Phys.\ B {\bf 600} (2001) 334
[hep-ph/0010150].



\bibitem{Anikin:2001ge}
I.~V.~Anikin and O.~V.~Teryaev,
Phys.\ Lett.\ B {\bf 509} (2001) 95
[hep-ph/0102209].

\bibitem{Radyushkin:2001fc}
A.~V.~Radyushkin and C.~Weiss,
hep-ph/0106059.

\bibitem{VGG}
M. Vanderhaeghen, P.A.M. Guichon, M. Guidal Phys.
Rev. Lett. {\bf 80} (1998) 5064.

\bibitem{BelMu}
A.~V.~Belitsky and D.~Muller,
Phys.\ Lett.\ B {\bf 486} (2000) 369
[hep-ph/0005028].

\bibitem{JiH} P.~Hoodbhoy and X.~Ji,
Phys.\ Rev.\ {\bf D 58} (1998) 054006
[hep-ph/9801369].

\bibitem{KB}
V.~M.~Braun and N.~Kivel,
Phys.\ Lett.\ B {\bf 501}, 48 (2001)
[hep-ph/0012220].

\bibitem{g2izmer}
G.~S.~Mitchell  [E155 Collaboration], ``Spin structure functions
$g_1$ and $g_2$ for the proton and deuteron,'' hep-ex/9903055;\\
P.~Bosted  [E155x Collaboration], Nucl.\ Phys.\  {\bf A663} (2000)
297.

\bibitem{Diehl:1997bu}
M.~Diehl, T.~Gousset, B.~Pire and J.~P.~Ralston,
Phys.\ Lett.\ B {\bf 411} (1997) 193
[hep-ph/9706344].

\bibitem{Diehl}
M.~Diehl,
Eur.\ Phys.\ J.\ C {\bf 19} (2001) 485
[hep-ph/0101335].


\bibitem{Anikin}
I.~V.~Anikin, B.~Pire and O.~V.~Teryaev,
Phys.\ Rev.\  {\bf D62} (2000) 071501 [hep-ph/0003203].

\bibitem{pion}
L.~L.~Frankfurt, M.~V.~Polyakov and M.~Strikman, hep-ph/9808449;\\
L.~Mankiewicz, G.~Piller and A.~Radyushkin, Eur.\ Phys.\ J.\  {\bf
C10} (1999) 307 [hep-ph/9812467];\\
 L.~L.~Frankfurt,
P.~V.~Pobylitsa, M.~V.~Polyakov and M.~Strikman, Phys.\ Rev.\
{\bf D60} (1999) 014010 [hep-ph/9901429];\\
M.~Penttinen,
M.~V.~Polyakov and K.~Goeke, Phys.\ Rev.\  {\bf D62} (2000) 014024
[hep-ph/9909489].

\bibitem{Dterm}
M.V. Polyakov and C. Weiss, Phys. Rev. {\bf D60} (1999) 114017.

\bibitem{RadDD}
A.~V.~Radyushkin, Phys.\ Rev.\  {\bf D59} (1999) 014030
[hep-ph/9805342].

\bibitem{GPV}
K.~Goeke, M.~V.~Polyakov and M.~Vanderhaeghen,
hep-ph/0106012.

\bibitem{bor}
V.~Y.~Petrov, P.~V.~Pobylitsa, M.~V.~Polyakov, I.~Bornig, K.~Goeke
and C.~Weiss, Phys.\ Rev.\  {\bf D57} (1998) 4325
[hep-ph/9710270].

\bibitem{KPV}
N.~Kivel, M.~V.~Polyakov and M.~Vanderhaeghen,
Phys.\ Rev.\ D {\bf 63} (2001) 114014
[hep-ph/0012136].

\bibitem{Spin} X.~Ji,
Phys.\ Rev.\ Lett.\  {\bf 78} (1997) 610
[hep-ph/9603249].

\end{thebibliography}
\end{document}